\documentclass[twoside,english,american]{IEEEtran}
\usepackage[T1]{fontenc}
\usepackage[latin9]{inputenc}
\usepackage{units}
\usepackage{textcomp}
\usepackage{amsmath}
\usepackage{amssymb}
\usepackage{graphicx}

\makeatletter

\providecommand{\tabularnewline}{\\}

\usepackage{cite}
\renewcommand{\citedash}{\citeright\hbox{--}\penalty\@m\citeleft}
\renewcommand{\citepunct}{\citeright,\penalty\@m\hskip.13emplus.1emminus.1em\citeleft}

\DeclareMathOperator{\rect}{\mathrm{rect}}

\makeatother

\usepackage{babel}
\begin{document}

\title{Optical Time-Frequency Packing: Principles, Design, Implementation,
and Experimental Demonstration}

\author{Marco Secondini \IEEEmembership{Member, IEEE}, Tommaso Foggi, Francesco
Fresi, Gianluca Meloni, Fabio Cavaliere \IEEEmembership{Member, IEEE},
Giulio Colavolpe \IEEEmembership{Senior Member, IEEE}, Enrico Forestieri
\IEEEmembership{Member, IEEE}, Luca Potì \IEEEmembership{Member, IEEE},
Roberto Sabella \IEEEmembership{Senior Member, IEEE}, and Giancarlo
Prati \IEEEmembership{Fellow, IEEE}%
\thanks{This work was supported in part by the Italian Ministry for Education
University and Research (MIUR) under the FIRB project COTONE.%
}%
\thanks{M. Secondini, F. Fresi, E. Forestieri, and G. Prati are with TeCIP
Institute, Scuola Superiore Sant'Anna, 56124 Pisa, Italy, and with
National Laboratory of Photonic Networks, CNIT, 56124 Pisa, Italy
(email: marco.secondini@sssup.it; francesco.fresi@sssup.it; forestieri@sssup.it;
giancarlo.prati@cnit.it).%
}%
\thanks{T. Foggi is with CNIT, 43124 Parma, Italy (email: tommaso.foggi@cnit.it).%
}%
\thanks{G. Meloni and L. Potì are with National Laboratory of Photonic Networks,
CNIT, 56124 Pisa, Italy (email: gianluca.meloni@cnit.it; luca.poti@cnit.it).%
}%
\thanks{F. Cavaliere and R. Sabella are with Ericsson (email: fabio.cavaliere@ericsson.com;
roberto.sabella@ericsson.com).%
}%
\thanks{G. Colavolpe is with the Dipartimento di Ingegneria dell'Informazione,
University of Parma, 43124 Parma, Italy (email: giulio@unipr.it).%
}%
\thanks{Copyright (c) 2015 IEEE. Personal use of this material is permitted.
However, permission to use this material for any other purposes must
be obtained from the IEEE by sending a request to pubs-permissions@ieee.org.%
}}
\maketitle
\begin{abstract}
Time-frequency packing (TFP) transmission provides the highest achievable
spectral efficiency with a constrained symbol alphabet and detector
complexity. In this work, the application of the TFP technique to
fiber-optic systems is investigated and experimentally demonstrated.
The main theoretical aspects, design guidelines, and implementation
issues are discussed, focusing on those aspects which are peculiar
to TFP systems. In particular, adaptive compensation of propagation
impairments, matched filtering, and maximum a posteriori probability
detection are obtained by a combination of a butterfly equalizer and
four 8-state parallel Bahl-Cocke-Jelinek-Raviv (BCJR) detectors. A
novel algorithm that ensures adaptive equalization, channel estimation,
and a proper distribution of tasks between the equalizer and BCJR
detectors is proposed. A set of irregular low-density parity-check
codes with different rates is designed to operate at low error rates
and approach the spectral efficiency limit achievable by TFP at different
signal-to-noise ratios. An experimental demonstration of the designed
system is finally provided with five dual-polarization QPSK-modulated
optical carriers, densely packed in a 100\,GHz bandwidth, employing
a recirculating loop to test the performance of the system at different
transmission distances.\end{abstract}

\section{Introduction}

Next generation optical systems will use coherent detection and advanced
signal processing for enabling the transmission of extremely high
bit rates. Currently deployed 100~Gb/s single-carrier systems typically
operate on a 50~GHz grid spacing, employing quadrature phase-shift
keying (QPSK) modulation with polarization multiplexing to meet the
required 2~bit/s/Hz spectral efficiency (SE), with a potential reach
of thousands of kilometers. This, considering the actual power and
bandwidth limitations of typical fiber-optic links, still leaves a
significant margin for improvement with respect to channel capacity.
On the other hand, accommodating the ever increasing traffic demand
will require, in the next years, to operate as close as possible to
the Shannon limit, achieving a much higher SE and possibly adapting
it to the available signal-to-noise ratio (SNR). While the use of
coherent detection, digital signal processing (DSP), and soft-decision
forward error correction is not in question in long-haul systems,
a few different options are being considered for the optical transport
format. Besides a high SE, the selected format should also offer best
performance in terms of energy efficiency, cost, and reliability,
the complexity of the required DSP being one of the driving factors
for all those issues. From a system point of view, the whole problem
can be summarized as finding the best combination of modulation and
coding that maximizes SE for a given SNR and constrained complexity.
In optical communications, orthogonal signaling is typically adopted
to ensure the absence of inter-symbol interference (ISI) and inter-carrier
interference (ICI). For instance, both Nyquist wavelength-division
multiplexing (WDM) \cite{JLT-1101-Bos} and orthogonal frequency-division
multiplexing (OFDM) \cite{Shieh:JOPNET08} solutions, whose performance
and complexity are basically equivalent on the fiber-optic channel
\cite{BaCoFo10}, employ orthogonal signaling. In both cases, the
orthogonality condition sets a lower limit to time- and frequency-spacing
(the Nyquist criterion), such that the achievable SE is limited by
the number of levels of the underlying modulation format. In fact,
higher SE requires higher-order modulation (e.g., 16-ary quadrature
amplitude modulation (QAM)), with higher transmitter and receiver
complexity.

Recently, a different approach has been proposed which, giving up
the orthogonality condition, allows to overcome the Nyquist limit
and achieve a higher SE with low-order modulations \cite{BaFeCo09b,Colavolpe11:opex,Poti:ECOC12,Sambo2014_JLT_TFP,Colavolpe2014_TCOM_TFPacking}.
This time- and frequency-packing (TFP) approach is an extension of
well known faster-than-Nyquist (FTN) signaling \cite{Ma75c}. In Mazo
FTN signaling, pulses are packed closer than the Nyquist limit without
asymptotic performance degradation, provided that the minimum Euclidean
distance of the system is not reduced and the optimum detector is
employed (Mazo limit) \cite{Ma75c}. Analogously, a closer packing
can be achieved also in frequency domain without asymptotic performance
degradation (two dimensional Mazo limit) \cite{RuAn05,Rusek09,Anderson13,Dasalukunte14}.
In other words, by increasing signaling rate for a fixed pulse bandwidth
(or, equivalently, by reducing pulse bandwidth for a fixed signaling
rate), some bandwidth resources are saved at the expense of introducing
ISI. A similar approach has been experimentally demonstrated also
in \cite{Cai12}. FTN%
\footnote{Here and in the rest of the paper, FTN refers to Mazo FTN, as originally
proposed in \cite{Ma75c}.%
}, however, does not provide the best performance in terms of SE, and
has a limit in the complexity of the required detector (which can
be very high). On the other hand, TFP overcomes this limit and seeks
the best solution by dividing the problem in three parts: i) set the
desired input constellation (e.g., QPSK) and detector complexity;
ii) find the optimum time- and frequency-spacing which provide the
maximum achievable SE for the given input constellation and detector
complexity; iii) select a proper code to approach as close as desired
the achievable SE (information theory guarantees that such a code
exists).

In this work, after introducing the theoretical aspects of the TFP
approach, we discuss the design procedure and implementation of a
TFP fiber-optic system and experimentally investigate its performance.
Section~\ref{sec:Time-Frequency-Packing} introduces the TFP approach
and the basic concept of achievable SE for a mismatched decoder, whose
maximization is the key aspect of TFP. In Section~\ref{sec:System-Design}
we explain how to design a TFP system and find the optimum modulation
parameters (time and frequency spacing) that maximize the achievable
SE for a given transmitter and receiver complexity. Moreover, we present
a set of irregular low-density parity-check (LDPC) codes to practically
approach the achievable SE. The practical implementation of a TFP
fiber-optic system is discussed in Section~\ref{sec:System-Implementation},
focusing on the DSP part, which is the only one to require some modifications
with respect to a standard WDM system employing coherent detection.
The experimental demonstration of the designed TFP system is addressed
in Section~\ref{sec:Experimental-Demonstration}: five closely-packed
40\,GBd dual-polarization (DP) quaternary phase-shift keying (QPSK)
channels are transmitted through a recirculating loop, keeping the
net SE beyond the theoretical limit of Nyquist-WDM (4\,bit/s/Hz)
up to 6000\,km; higher SEs are achieved at shorter distances by adapting
the TFP configuration and code rate to the available OSNR, achieving
a net SE of more than 7\,bit/s/Hz (for DP-QPSK transmission) at a
distance of 400\,km. A discussion of the results is provided in Section\,\ref{sec:Discussion}
and conclusions are finally drawn in Section\,\ref{sec:Conclusions}.

\section{Time-Frequency Packing\label{sec:Time-Frequency-Packing}}

\begin{figure}
\begin{centering}
\includegraphics[width=1\columnwidth]{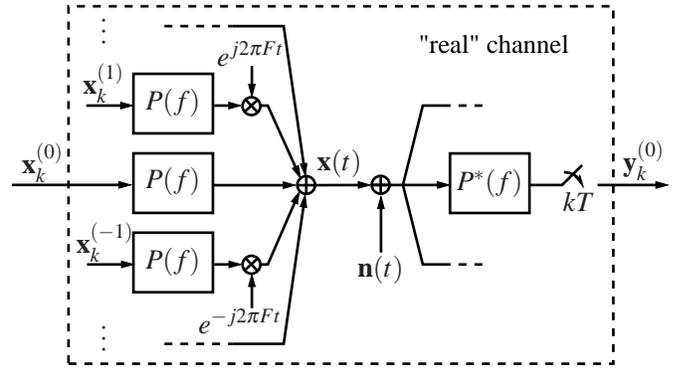}
\par\end{centering}

\caption{\label{fig01}Low-pass equivalent model employed to design the TFP
system}
\end{figure}

In order to summarize the general ideas behind TFP, we refer here
to an ideal dual-polarization AWGN channel. Rather than as a specific
modulation format, TFP should be regarded as a design procedure for
the optimization of a class of modulation formats---namely, multicarrier
linear modulations, to which both Nyquist WDM and OFDM belong. Many
communication systems employ this kind of modulation to encode information
onto waveforms which can be practically generated and reliably transmitted
through a given communication channel. The low-pass equivalent model
of a generic linearly-modulated multicarrier system is schematically
depicted in Fig.~\ref{fig01}. 
All the equally-spaced carriers are linearly modulated with the same
modulation format and shaping pulse $p(t)=\mathcal{F}^{-1}\{P(f)\}$.
The complex envelope of the transmitted signal is 
\begin{equation}
\mathbf{x}(t)=\sum_{\ell=-M}^{M}\sum_{k=1}^{K}\mathbf{x}_{k}^{(\ell)}p(t-kT)e^{j2\pi\ell Ft}\label{eq:TFP signal}
\end{equation}
where $\mathbf{x}_{k}^{(\ell)}$ is the transmitted symbol (a two-component
vector, one per each polarization) on the $\ell$-th carrier at time
$kT$, $T$ is the symbol time (or time spacing between adjacent symbols),
$F$ the frequency spacing between adjacent carriers, and, for simplicity,
a perfect time and phase synchronization among the carriers is assumed%
\footnote{When employing single-user detectors, which is the case considered
in this work, the actual phase and time shift between carriers is
typically irrelevant and has a negligible impact on the achievable
information rate, both in terms of linear cross-talk among carriers
(which, as shown later, are only slightly overlapped), and of inter-channel
nonlinearity \cite{Secondini:JLT2013-AIR}. In fact, the experimental
demonstration of Section~\ref{sec:Experimental-Demonstration} does
not employ any phase locking or time synchronization among carriers.%
}. Signal (\ref{eq:TFP signal}) is corrupted by additive white Gaussian
noise (AWGN) $\mathbf{n}(t)$ and demodulated by a bank of matched
filters and symbol-time samplers. Assuming a system with $2M+1$ carriers
(a number sufficiently large to neglect border effects) and denoting
by $\mathbf{x}=\{\mathbf{x}_{k}^{(\ell)}\}$ the set of transmitted
symbols and by $\mathbf{y}=\{\mathbf{y}_{k}^{(\ell)}\}$ the set of
channel outputs, the SE of the system (bit/s/Hz) is%
\footnote{Here, $\eta$ denotes the maximum rate per unit bandwidth at which
information can be reliably transmitted through the channel with i.i.d.
input symbols, where maximization is performed over all possible encoding
of information bits on transmitted symbols. In the following, we will
also consider other two slightly different definitions of SE, where
specific constraints on the detector or coding strategy are imposed.
Moreover, upper-case letters denote random variables, while lower-case
letters denote their realizations.%
}
\begin{equation}
\eta=\frac{I(\mathbf{X};\mathbf{Y})}{FT}\label{eq:eff.spettrale}
\end{equation}
where $I(\mathbf{X};\mathbf{Y})$ is the average mutual information
rate (bit/symbol) between input and output \cite{Gallager68}, and
$1/FT$ is the inverse of the time-frequency spacing product, which
equals the number of symbols transmitted per second per Hertz. Typically,
these modulation formats are designed to avoid both ISI and ICI by
imposing proper orthogonality constraints on the employed waveforms
(e.g., Nyquist-WDM or OFDM). This, in turn, poses a constraint on
the pulse shape and sets a limit to the minimum time and frequency
spacing between pulses (Nyquist limit). When this orthogonal signaling
approach is employed, $I(\mathbf{X};\mathbf{Y})=I(\mathbf{X}_{k}^{(\ell)};\mathbf{Y}_{k}^{(\ell)}),\,\forall k,\ell$
is achievable by a symbol-by-symbol detector%
\footnote{Note that with symbol-by-symbol detector we mean a simple threshold
detector not working on a trellis. So, according to our definition,
a BCJR detector is not a symbol-by-symbol detector.%
} and depends only on the modulation format and signal-to-noise ratio.
On the other hand, the minimum value of the time-frequency spacing
product at denominator of (\ref{eq:eff.spettrale}) is set by the
Nyquist limit and is $FT=1$---achievable, for instance, by using
pulses with a rectangular spectrum $P(f)=\sqrt{1/B}\rect(f/B)$ of
low-pass bandwidth $B/2$ and setting $F=B$ and $T=1/B$, as in Nyquist-WDM
\cite{JLT-1101-Bos}. Thus, the only way to increase $\eta$ is through
the numerator of (\ref{eq:eff.spettrale}), by increasing the cardinality
of the modulation alphabet. The main drawback of this approach is
that a high SE is obtained at the expense of strict requirements on
spectral shaping and higher-order symbol alphabets. For orthogonal
DP-QPSK modulation, the upper limit is $\eta=\unit[4]{bit/s/Hz}$.

On the contrary, if we give up the orthogonality condition, we have
no constraints on the choice of $p(t)$, $F$, and $T$. Thus, we
can select a shaping pulse $p(t)$ that is compatible with the available
hardware components, and try to increase (\ref{eq:eff.spettrale})
by reducing the denominator $FT$ below the Nyquist limit, without
changing the symbol alphabet. This way, however, we also introduce
ICI and ISI and, therefore, reduce the numerator $I(\mathbf{X};\mathbf{Y})$.
Thus, we can transmit more symbols per unit time and frequency, but
less information bits per symbol. Moreover, the presence of ICI and
ISI makes $I(\mathbf{X};\mathbf{Y})$ unachievable by a symbol-by-symbol
detector. Thus, the problem is that of selecting $F$ and $T$ such
that the decrease of the numerator of (\ref{eq:eff.spettrale}) is
more than balanced by the decrease of the denominator, for any allotted
complexity of the required detector. In Mazo FTN signaling, $T$ is
selected as the minimum value for which the minimum Euclidean distance
of the system is not reduced (Mazo limit) \cite{Ma75c}. Thus, pulses
are packed closer than the Nyquist limit without performance degradation,
provided that the optimum sequence detector is employed. Analogously,
a closer packing can be achieved also in frequency domain without
performance degradation (two-dimensional Mazo limit) \cite{RuAn05,Rusek09,Anderson13,Dasalukunte14}.
This approach, however, does not provide the best performance in terms
of (\ref{eq:eff.spettrale}) and poses no constraints on the complexity
of the required detector.

The TFP approach, instead, allows to introduce, for a given constellation,
an arbitrary constraint on the detector complexity (meant as specified
later) and optimize the time and frequency spacings in order to maximize
the spectral efficiency achievable by that constrained-complexity
detector when using that constellation. To this aim---limiting the
analysis to single-user detectors (which is a practical choice to
limit the detector complexity) and focusing, without loss of generality,
on the central carrier ($\ell=0$), such that the others are considered
only as a source of ICI and are, therefore, a part of the ``real''
channel depicted in Fig.~\ref{fig01}---a slightly different (and
more practical, as shown later) definition of SE is adopted by replacing
the mutual information rate $I(\mathbf{X};\mathbf{Y})$ in (\ref{eq:eff.spettrale})
with the achievable information rate (AIR) for a mismatched decoder~\cite{MeKaLaSh94}
\begin{equation}
\hat{I}(\mathbf{X}^{(0)};\mathbf{Y}^{(0)})\triangleq\lim_{K\rightarrow\infty}\frac{1}{K}E\Bigg\{\log\frac{q(\mathbf{y}^{(0)}\arrowvert\mathbf{x}^{(0)})}{q(\mathbf{y}^{(0)})}\Bigg\}\le I(\mathbf{X};\mathbf{Y})\label{eq:AIR}
\end{equation}
where vectors $\mathbf{x}^{(0)}$ and $\mathbf{y}^{(0)}$ collect,
respectively, the $K$ transmitted symbols and $K$ received samples
on the ``real'' channel depicted in Fig.\,\ref{fig01}, expectation
$E\{\cdot\}$ is taken with respect to the real channel, while $q(\mathbf{y}^{(0)}\arrowvert\mathbf{x}^{(0)})$
and $q(\mathbf{y}^{(0)})$ are, respectively, the conditional and
marginal output distribution obtained by connecting inputs to an arbitrary
auxiliary channel---equality holding if $q(\mathbf{y}^{(0)}\arrowvert\mathbf{x}^{(0)})$
and $q(\mathbf{y}^{(0)})$ equal the distributions of the real channel.
The importance of the quantity defined in (\ref{eq:AIR}) is in its
properties, which hold for any real and auxiliary channel: it is a
lower bound to the mutual information rate $I(\mathbf{X};\mathbf{Y})$
on the real channel; it is achievable by the maximum \textit{a posteriori}
probability (MAP) detector designed for the selected auxiliary channel;
and it can be simply evaluated through simulations \cite{ArLoVoKaZe06}.
The auxiliary channel, though arbitrary, is conveniently chosen as
the one providing the best trade-off between performance and complexity:
the closer the auxiliary channel to the real channel, the higher is
the AIR $\hat{I}(\mathbf{X}^{(0)};\mathbf{Y}^{(0)})$ (and closer
to $I(\mathbf{X};\mathbf{Y})$); the simpler the auxiliary channel,
the simpler is the MAP detector required to achieve $\hat{I}(\mathbf{X}^{(0)};\mathbf{Y}^{(0)})$.
Finally, time and frequency spacing are optimized by maximizing the
achievable SE with the selected detector
\begin{equation}
\hat{\eta}_{\mathrm{max}}=\max_{F,T>0}\frac{\hat{I}(\mathbf{X}^{(0)};\mathbf{Y}^{(0)})}{FT}\le\eta.\label{eq:max.ach.spettr.eff.}
\end{equation}
Optimization (\ref{eq:max.ach.spettr.eff.}) is the very essence of
TFP, which distinguishes it from FTN or other non-orthogonal signaling
techniques. The result obtained through (\ref{eq:max.ach.spettr.eff.})
depends on the shaping pulse $p(t)$ considered in (\ref{eq:TFP signal}),
though any scaling of the pulse in time domain can be easily accounted
for by properly rescaling $T$ and $F$. The maximum achievable SE
depends also on the given SNR (it increases as the SNR increases).
However, the optimum $F$ and $T$ depend only slightly on it, such
that a single optimization can be adopted for a wide range of SNRs
(i.e., of link distances).

The last step of the TFP method is common to almost any digital communication
system and consists in finding a coding strategy that, by properly
encoding information bits on transmitted symbols $\{\mathbf{x}_{k}^{(0)}\}$,
operates arbitrarily closely to (\ref{eq:max.ach.spettr.eff.})---information
theory guarantees that such a code does exist. Though similar coding
strategies can be adopted in TFP and orthogonal signaling, this step
has some peculiarities related to the presence of ISI and ICI which
will be discussed in the next section.

\section{System Design\label{sec:System-Design}}

In this section, we show how to design a multicarrier fiber-optic
system by employing the TFP approach described in the previous section.
We refer, again, to the ideal low-pass equivalent scheme reported
in Fig.~\ref{fig01}, which, under some assumptions, is a reasonable
representation of the fiber-optic channel,%
\footnote{This means that the TFP system is optimized in the back-to-back configuration.
However, in a practical implementation, most impairments (e.g, chromatic
and polarization mode dispersion) are compensated for by DSP, as explained
in Section\,\ref{sec:System-Implementation}, and do not change the
nature of the channel. The other impairments, such as fiber nonlinearity,
are responsible for a decrease of the AIR with respect to the AWGN
channel. These impairments can be simply included in the computation
of the achievable SE and in the optimization of the TFP system, as
briefly discussed later and detailed in \cite{Colavolpe2014_TCOM_TFPacking}.%
} and consider a DP-QPSK modulation alphabet. A sequence of i.i.d.
symbols $\{\mathbf{x}_{k}\}$, drawn from a DP-QPSK alphabet, modulates
the selected carrier ($\ell=0$, in the scheme) at rate $1/T$ with
a real shaping pulse $p(t)$. All the modulated carriers are then
combined and transmitted through an AWGN channel with noise $\mathbf{n}(t)$.
At the receiver, the selected carrier is demodulated by a matched
filter and a symbol-time sampler. Received samples $\{\mathbf{y}_{k}\}$
are finally sent to a MAP symbol detector that operates on the output
of the matched filter \cite{Ung:TCOM74,Colavolpe2005_ComLett_MAP-Ungerboeck}
and is implemented through the algorithm by Bahl, Cocke, Jelinek,
and Raviv (BCJR) \cite{BCJR}. The MAP symbol detector is matched
to an auxiliary channel, whose selection determines the distributions
$q(\mathbf{y}^{(0)}\arrowvert\mathbf{x}^{(0)})$ and $q(\mathbf{y}^{(0)})$
to be used in (\ref{eq:AIR}). In particular, as an auxiliary channel
we take an approximation of the real channel, obtained from the latter
by neglecting ICI, truncating ISI to the first $L_{T}\le L$ pre-
and post-cursor symbols (after the matched filter)---where $2L+1$
is the actual memory of the channel and $L_{T}$ is a design parameter
strictly related to detector complexity---and increasing the variance
of the noise samples at the output of the matched filter from the
actual value $\sigma_{n}^{2}$ to a numerically optimized value $\sigma_{n'}^{2}>\sigma_{n}^{2}$
to account for the neglected ICI and ISI. This choice provides a reasonable
trade-off between performance (how tight is the bound in (\ref{eq:AIR}))
and complexity (of the matched BCJR detector). Moreover, when a Gray-mapping
is used, the DP-QPSK modulation can be seen as the combination of
four orthogonal BPSK modulations, one per each quadrature component
of each state of polarization of the signal. Thus, four independent
and identical BCJR detectors with \foreignlanguage{english}{$2^{L_{T}}$}
states are used to separately detect the four BPSK components that
are transmitted over four independent and identical channels with
memory $2L_{T}+1$. 

\begin{figure*}
\begin{centering}
\includegraphics[width=0.5\textwidth]{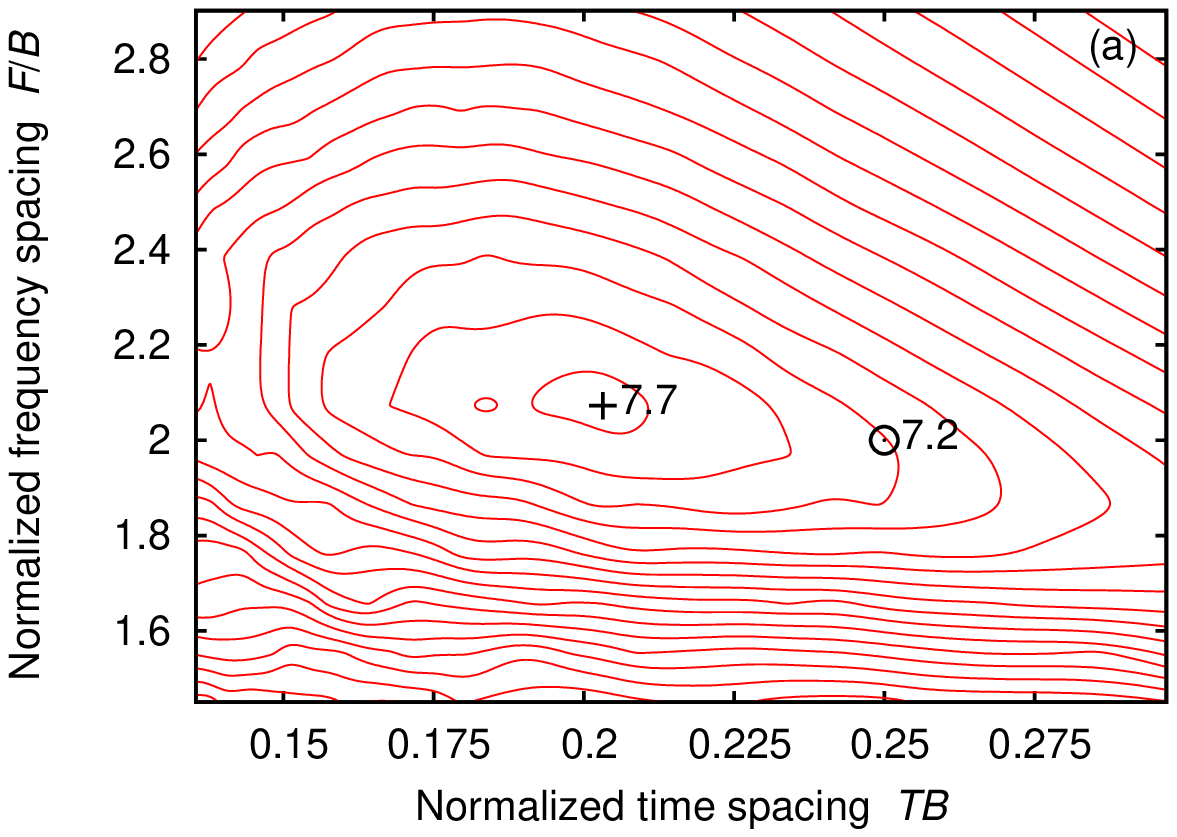}\includegraphics[width=0.5\textwidth]{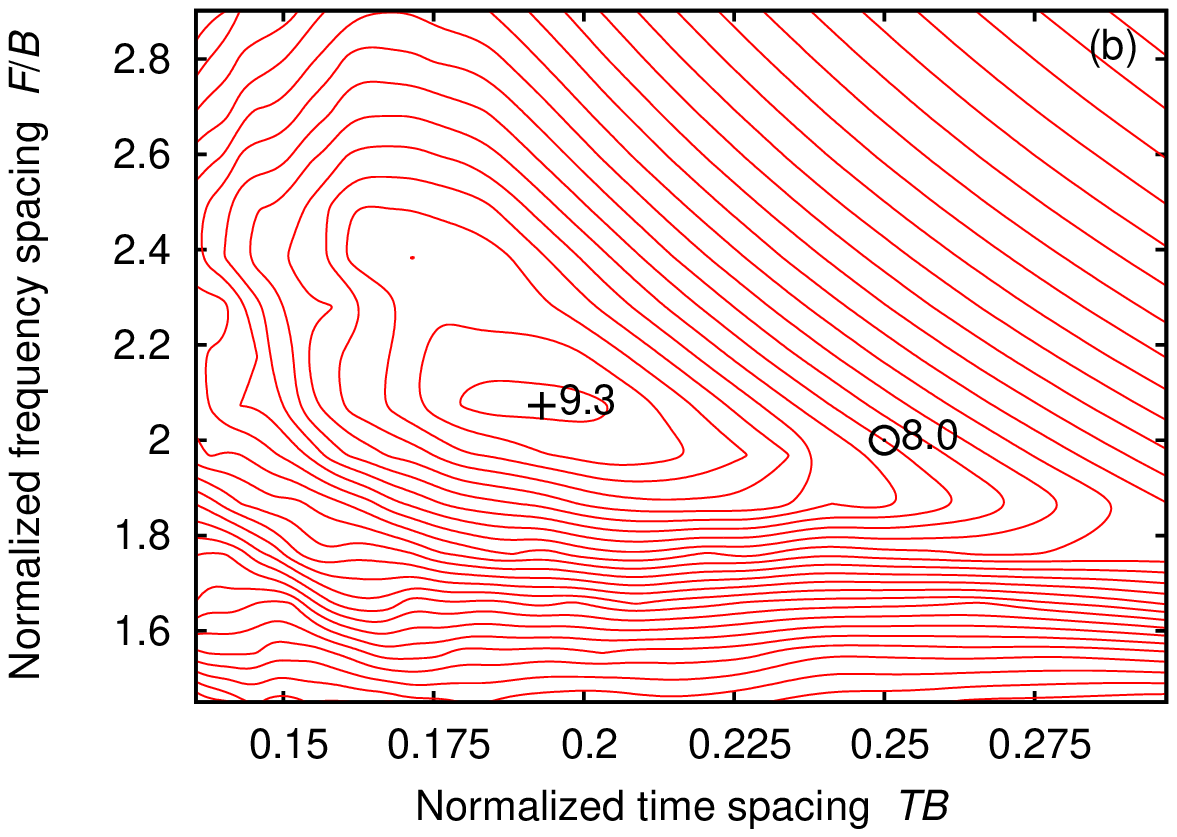}
\par\end{centering}

\caption{\label{fig02}Contour plots of the achievable SE (obtained by numerical
simulations and shown with increments of 0.2) as a function of the
normalized time and frequency spacing for DP-QPSK modulation on the
AWGN channel for: (a) $E_{b}/N_{0}=\unit[7.5]{dB}$; (b) $E_{b}/N_{0}=\unit[22.5]{dB}$.
The maximum value (+) and the value obtained with one of the configurations
adopted in the experimental setup ($\odot$) are also reported at
the corresponding coordinates.}
\end{figure*}

Although also the pulse shape $p(t)$ can be optimized to maximize
the SE \cite{Modenini:TCOM13-optimal_pulse}, in this work we consider
only the pulse shape obtained by employing a ninth-order type I Chebyshev
filter with 3\,dB bandwidth $B$, which approximately models the
electrical low-pass filters available in our laboratory for the experimental
implementation described in Section~\ref{sec:Experimental-Demonstration}.
For the given shape, (\ref{eq:AIR}) is evaluated through numerical
simulations as explained in \cite{ArLoVoKaZe06} on a grid of values
of the normalized time and frequency spacing $TB$ and $F/B$, seeking
the maximum SE (\ref{eq:max.ach.spettr.eff.}) and the corresponding
optimum spacings. To account for unsynchronized channels and unlocked
lasers, each modulated carrier is also subject to a random phase and
time shift and polarization rotation. The optimization is performed
considering a truncated channel memory $2L_{T}+1=7$ (requiring 8-state
MAP symbol detectors) and two different values of the SNR per bit
(defined as the ratio $E_{b}/N_{0}$ between the mean energy per bit
and the noise power spectral density and related to the OSNR through
$\mathrm{OSNR}=R_{b}E_{b}/(2N_{0}B_{\mathrm{ref}})$, where $R_{b}$
is the total net bit rate and $B_{\mathrm{ref}}\simeq\unit[12.5]{GHz}$
the conventional reference bandwidth) of 7.5 and 22.5\,dB. The corresponding
contour plots of the achievable SE are reported in Fig.~\ref{fig02}(a)
and (b).
 It can be noticed that the optimum configuration is obtained with
a significant overlap of pulses in time ($TB\simeq0.2)$) and only
a moderate overlap in frequency ($F/B\simeq2$). This is due to the
choice of a single-user detector (which cannot cope with ICI). On
the other hand, a multi-user detector would allow for a more relevant
packing also in frequency domain, providing a higher SE at the expense
of a higher complexity.

In principle, the TFP optimization procedure described here can be
applied also to a realistic fiber-optic channel, as the AIR definition
(\ref{eq:AIR}), its properties, and the simulation-based method for
its computation \cite{ArLoVoKaZe06} are valid for \emph{any} channel.
The only requirement is that of computing the output sequence $\{\mathbf{y}_{k}\}$
for the desired real channel (e.g., through the split-step Fourier
method). This, however, significantly increases the computation time
required to estimate a single AIR value and makes the optimization
procedure cumbersome. For this reason, we decided to optimize the
system in the absence of nonlinear effects, and then tested the obtained
suboptimum configuration (the one in Fig.\,\ref{fig02}) over a realistic
link. A numerical estimate of the achievable SE with the suboptimum
configuration over the realistic link (including nonlinear effects)
is reported in Section\,\ref{sec:Discussion} (Fig.\,\ref{fig:confronto-experimento-simulazioni})
and compared to experimental results. 

The achievable SE obtained with this design procedure can be practically
approached by employing properly designed codes. When the TFP technique
is adopted, and thus ISI is intentionally introduced, codes designed
for the AWGN channel no longer perform satisfactorily. So a redesign
is required. We designed proper LDPC codes specifically tailored for
the ISI channels resulting from the adoption of the TFP technique.
The adopted procedure is based on two steps. The heuristic technique
for the optimization of the degree distributions of the LDPC variable
and check nodes proposed in \cite{teKrAs04} is first adopted. This
technique consists in a curve fitting on extrinsic information transfer
(EXIT) charts, is based on a Gaussian assumption on all messages involved
in the iterative process, and is much simpler than other optimization
techniques, such as density evolution, which require intensive computational
efforts. The parameters of the designed codes are reported in Table
\ref{tab:parameters-of-the} where $r$ denotes the rate of the code
and the degree distributions of variable and check nodes are provided
by giving the fraction $a_{i}$ ($\sum_{i}a_{i}=1$) of degree $i$
nodes. In any case, the codeword length is $N=64800$. 
\begin{table*}
\caption{Code rates and degree distributions of the designed LDPC codes.\label{tab:parameters-of-the}}

\centering{}%
\begin{tabular}{|c|c|c|}
\hline 
$r$ & variable node distribution & check node distribution\tabularnewline
\hline 
\hline 
2/3 & {\scriptsize{$\begin{array}{ccc}
a_{2}=0.333318 & a_{3}=0.6 & a_{13}=0.0666821\end{array}$}} & {\scriptsize{$\begin{array}{ccc}
a_{9}=0.000277778 & a_{10}=0.998935 & a_{11}=0.000787037\end{array}$}}\tabularnewline
\hline 
3/4 & {\scriptsize{$\begin{array}{ccc}
a_{2}=0.249985 & a_{3}=0.666682 & a_{12}=0.0833333\end{array}$}} & {\scriptsize{$\begin{array}{ccc}
a_{13}=0.000679012 & a_{14}=0.99858 & a_{15}=0.000740741\end{array}$}}\tabularnewline
\hline 
4/5 & {\scriptsize{$\begin{array}{ccc}
a_{2}=0.2 & a_{3}=0.699985 & a_{11}=0.100015\end{array}$}} & {\scriptsize{$\begin{array}{cc}
a_{18}=0.999383 & a_{19}=0.000617284\end{array}$}}\tabularnewline
\hline 
5/6 & {\scriptsize{$\begin{array}{cccc}
a_{1}=1.54321\cdot10^{-5} & a_{2}=0.166651 & a_{3}=0.75 & a_{13}=0.0833333\end{array}$}} & {\scriptsize{$\begin{array}{cc}
a_{21}=9.25926\cdot10^{-5} & a_{22}=0.999907\end{array}$}}\tabularnewline
\hline 
8/9 & {\scriptsize{$\begin{array}{ccc}
a_{2}=0.111096 & a_{3}=0.777793 & a_{4}=0.111111\end{array}$}} & {\scriptsize{$\begin{array}{cc}
a_{27}=0.999861 & a_{28}=0.000138889\end{array}$}}\tabularnewline
\hline 
\end{tabular}
\end{table*}

\begin{figure}
\begin{centering}
\includegraphics[width=1\columnwidth]{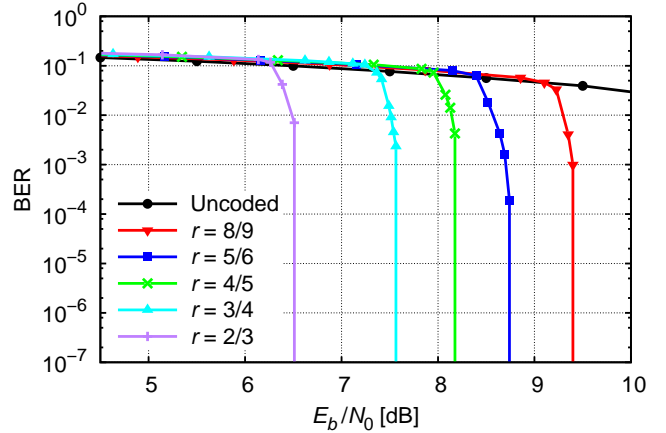}
\par\end{centering}

\caption{\label{fig03}Back-to-back BER for uncoded transmission and for the
designed LDPC codes, obtained through numerical simulations with the
TFP configuration adopted in the experimental setup (40\,GBd configuration).}
\end{figure}

Once the degree distributions of the LDPC variable and check nodes
have been designed, the parity check matrix of an LDPC code with those
degree distributions is built through the very effective PEG algorithm
\cite{XiBa04,HuElAr05}, which allows to design an LDPC code whose
underlying Tanner graph has a large girth. The BER curves for uncoded
QPSK transmission and for the designed LDPC codes (independent encoding
of the in-phase and quadrature components of each polarization) obtained
through numerical simulations for the back-to-back system with the
TFP configuration adopted in the experimental setup (constrained optimum
at 40\,GBd and frequency spacing $F=\unit[20]{GHz}$) are reported
in Fig.~\ref{fig03}. With this TFP configuration, the 8/9 LDPC code
requires $E_{b}/N_{0}\simeq\unit[9.3]{dB}$ and provides an SE of
about 7.1\,bit/s/Hz. Thus, we can compare it with Fig.\,\ref{fig02}(a),
which shows that approximately the same SE (7.2\,bit/s/Hz) is theoretically
achievable at $E_{b}/N_{0}=\unit[7.5]{dB}$. This means that the designed
code over the (back-to-back) TFP channel has a penalty of less than
2\,dB with respect to the theoretical limit provided by the AIR in
(\ref{eq:max.ach.spettr.eff.}). The gap between the actual rates
achieved by the designed codes over the fiber-optic channel (including
nonlinear effects) and the AIR over the same channel is numerically
investigated in Section\,\ref{sec:Discussion} (Fig.\,\ref{fig:confronto-experimento-simulazioni})
and is between 2 and 3\,dB for all the codes. Detection of this kind
of codes is typically characterized by the presence of error floors
at high SNRs. In our simulations, we transmitted up to 10000 codewords
without observing any floor, meaning that error floors, if present,
are probably located at a BER lower than $10^{-8}$. In any case,
outer hard-decision external codes with very low overhead can be employed
to correct the residual errors and remove the floor. For instance,
in the DVB-S2 standard, where LDPC codes with same length and rates
as in Tab.\,\ref{tab:parameters-of-the} are adopted, outer BCH codes
with less than 0.4\% overhead are used to correct from 8 to 12 residual
errors (depending on the rate) per codeword \cite{Standard-DVB-S2}.
In this case, an interleaver might be needed between the inner LDPC
and outer BCH code, increasing the latency of the system. 

\section{System Implementation\label{sec:System-Implementation}}

\begin{figure*}
\centering{}\includegraphics[width=1\textwidth]{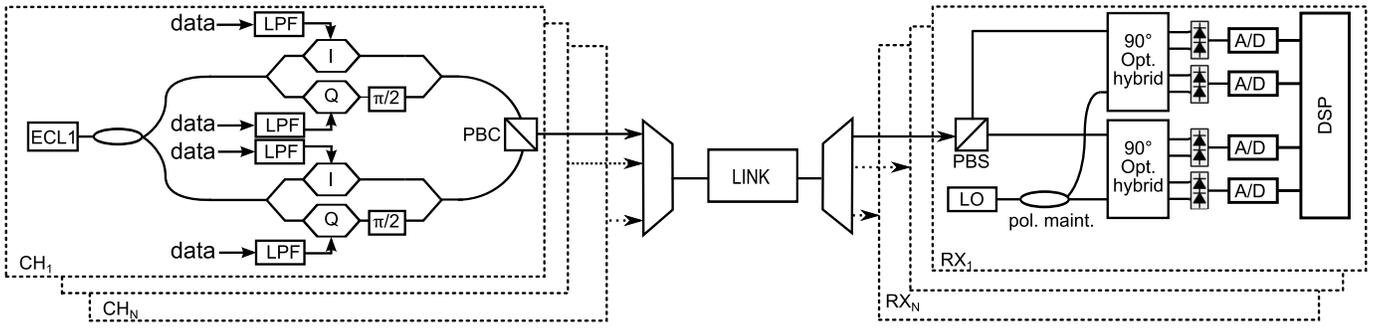}\caption{\label{fig04}Transmitter and receiver schemes of a TFP system employing
DP-QPSK modulation.}
\vspace*{3ex}
\end{figure*}

A fiber-optic system based on the TFP approach can be implemented
by using the same hardware configuration typically used for WDM systems
based on coherent detection \cite{Col:JLT0709}. A significant difference
is in the DSP algorithms actually required at the receiver. Moreover,
some care should be taken to ensure that the transmitted signal is
linearly modulated as in~(\ref{eq:TFP signal}). In this section,
we will refer to the transmitter and receiver implementation schemes
shown in Fig.~\ref{fig04}, focusing on those elements which are
peculiar to the TFP implementation. Practical details about the experimental
setup actually employed in the experimental demonstration will be
given in Section~\ref{sec:Experimental-Demonstration}.

Since the system employs single-user detectors, an independent transmitter
and receiver pair is used per each optical carrier. Each optical carrier
is thus generated at the desired wavelength (e.g., by an external-cavity
laser (ECL)), modulated, optically multiplexed with the other modulated
carriers, transmitted through the optical link, extracted by an optical
demultiplexer, and independently detected. In each transmitter, the
in-phase and quadrature components of two orthogonal states of polarization
are independently and linearly modulated by a pair of nested Mach-Zehnder
modulators (MZMs). In principle, the desired pulse shape $p(t)$ can
be obtained either operating on the electrical signals that drive
the modulator (through a low-pass filter (LPF), as actually shown
in the scheme of Fig.\,\ref{fig04}) or on the optical signal after
the modulator (through an optical band-pass filter), provided that
the overall equivalent low-pass impulse response of the transmitter
(driver, modulator, electrical filter and/or optical filter) is $p(t)$
and that linearity of the modulator is preserved by employing a driving
voltage significantly lower than the modulator half-wave voltage $V_{\pi}$.
A comparison of the performance obtained by optical or electrical
filtering when employing different driving voltages is presented in
\cite{Fresi:PTL2013_TFP_OEfilter}. An alternative modulation scheme,
where linear modulation (\ref{eq:TFP signal}) is obtained by operating
the MZM at its maximum driving voltage (to reduce its insertion loss),
may also be devised. For instance, by using an additional MZM as a
pulse carver and employing optical filtering to obtain the desired
pulse shape $p(t)$ and $BT$ product \cite{Colavolpe11:opex,Colavolpe2014_TCOM_TFPacking},
the nonlinearity of the MZM affects only (and slightly) the overall
pulse shape $p(t)$, but does not introduce nonlinear ISI. In this
scheme, however, the insertion loss saved by increasing the MZM driving
voltage is replaced by the additional loss introduced by the pulse
carver and optical filter. Finally, a possible implementation based
on an arrayed waveguide grating device that fi{}lters and multiplexes
all the frequency subchannels in the optical domain has been proposed
in \cite{Abrardo2014_JLT_TFP}. In this work, we consider a modulation
scheme based on a single MZM (driven at low voltage) and analogue
electrical filtering which, at the present, seems to be the most practical
choice in terms of cost and complexity. Moreover, as discussed in
Section~\ref{sec:System-Design}, the choice of $p(t)$ is not critical
and a reasonably good performance can be obtained by employing available
analogue low-pass filters, as shown in Fig.~\ref{fig02}.

\begin{figure}
\begin{centering}
\includegraphics[width=1\columnwidth]{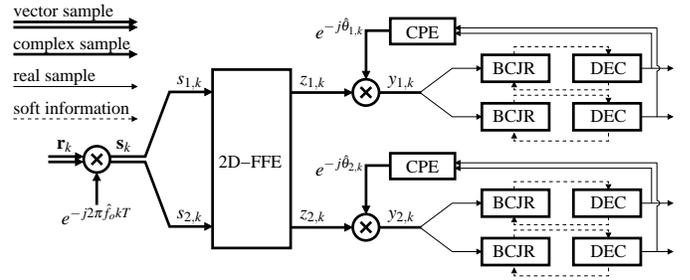}
\par\end{centering}

\caption{\label{fig05}Digital signal processing scheme.}
\end{figure}

At the receiver side, each optical carrier is demodulated by employing
a phase- and polarization-diversity coherent detection scheme. After
optical demultiplexing, each carrier is split into two orthogonal
states of polarizations, which are then separately combined with the
optical field of a local oscillator (LO) laser in a $2\times4$ 90\textdegree{}
optical hybrid and detected with two pairs of balanced photodetectors.
The four resulting electrical signals (the in-phase and quadrature
components of each state of polarization) are then sampled by an A/D
converter with a bandwidth $B_{s}\ge B$, where $B$ is the (low-pass)
bandwidth of the shaping pulse $p(t)$, and a sampling rate $R_{s}\ge2B_{s}$.
The remaining part of receiver processing is digitally implemented
according to the scheme depicted in Fig.~\ref{fig05}, assuming a
sampling rate $R_{s}=1/T$. Note that, since TFP is employed, the
required bandwidth and sampling rate are typically lower than $1/(2T)$
and $1/T$, respectively, and digital upsampling can be employed to
achieve the sampling rate $R_{s}=1/T$ required for symbol-time processing,
without any performance degradation.
 The $k$-th received column vector of samples $\mathbf{r}_{k}=(r_{1,k},r_{2,k})^{T}$
(one complex sample per state of polarization) is first processed
to compensate for the presence of any large and slowly varying frequency
offset $f_{o}$ between the transmit and receive lasers. The estimate
$\hat{f}_{o}$ is obtained during the training phase (on a known training
sequence) by employing the frequency estimation algorithm described
in \cite{Mengali97_TCOM_freq-est}, and then slowly updated based
on decisions. Compensated samples $\mathbf{s}_{k}=\mathbf{r}_{k}e^{-j2\pi\hat{f}_{0}kT}$
are then processed by an adaptive 2-D $N_{c}$-tap synchronous feed-forward
equalizer (FFE) that compensates for linear propagation impairments,
such as group-velocity dispersion (GVD), polarization rotations, and
polarization-mode dispersion (PMD), and completes (as explained later)
the implementation of the matched filter. Symbol timing synchronization
is also performed by the adaptive equalizer. As the system performance
is nearly independent of the clock phase when the period of the equalizer
transfer matrix ($1/T$ when a synchronous equalizer is employed)
is larger than twice the low-pass bandwidth of the signal (about $1/4T$
in our TFP implementation, as shown in Fig.~\ref{fig02}), a simple
feedback signal for the clock rate can be obtained by monitoring the
dynamics of the equalizer coefficients, which reflect the drift of
the clock phase due to an error of the clock rate \cite{Ung:TCOM76}.
However, in the experimental implementation of Section~\ref{sec:Experimental-Demonstration},
in which signal samples are collected in blocks of 1 million per quadrature,
the drift of the clock phase is negligible and no clock recovery is
employed. At the output of the equalizer, the components $z_{1,k}$
and $z_{2,k}$ of the equalized samples $\mathbf{z}_{k}$ are then
separated and independently processed. For each component, decision-directed
carrier phase estimation (CPE) based on the Tikhonov parametrization
algorithm \cite{Colavolpe2005_JSAC_phase-noise} is employed to cope
with the laser phase noise%
\footnote{The algorithm in \cite{Colavolpe2005_JSAC_phase-noise} is properly
modified to use (preliminary) hard decisions (which are a particular
case of soft decisions used in \cite{Colavolpe2005_JSAC_phase-noise})
and to account for ISI by filtering the preliminary decisions with
the overall impulse response of the channel, estimated as explained
later. The algorithm is also modified to extract hard (rather than
soft) estimates $\hat{\theta}_{1,k}$ and $\hat{\theta}_{2,k}$ of
the carrier phase for the two polarization components. Considering
the notation in \cite{Colavolpe2005_JSAC_phase-noise}, and in particular
the factor graph in Fig.~2 of \cite{Colavolpe2005_JSAC_phase-noise},
the product of the incoming messages $p_{d}(\theta_{k})$, $p_{f}(\theta_{k})$,
$p_{b}(\theta_{k})$ to variable node $\theta_{k}$ is proportional
to the a posteriori probability density function of the carrier phase
$\theta_{k}$ at discrete time $k$ given the received sequence. Its
argmax is thus the MAP estimate $\hat{\theta}_{k}$ of $\theta_{k}$
(for either of the two polarizations). In particular, having the a
posteriori probability density function of $\theta_{k}$ a Tikhonov
distribution, the MAP estimate we are looking for equals the argument
of the complex parameter of this distribution.%
}. Finally, the in-phase and quadrature components of the compensated
samples $y_{1,k}=z_{1,k}e^{-j\hat{\theta}_{1,k}}$ and $y_{2,k}=z_{2,k}e^{-j\hat{\theta}_{2,k}}$
are separated and sent to four parallel $2^{L_{T}}$-state BCJR detectors
\cite{BCJR}, followed by four LDPC decoders. The BCJR detectors and
LDPC decoders iteratively exchange information to achieve MAP detection
according to the turbo principle \cite{hagenauer1997turbo}. At each
iteration, as new (more accurate) preliminary decisions are available
from the decoders, the CPEs update the phase estimates $\hat{\theta}_{1,k}$
and $\hat{\theta}_{2,k}$ and a new set of compensated samples $y_{1,k}$
and $y_{2,k}$ is fed to the BCJR detectors. At the first iteration,
as preliminary decisions are not available, the CPEs exploit pilot
symbols (evenly inserted in the transmitted sequence at rate $r_{p}$)
to provide a rough initial estimate of the phase and make the iterative
process bootstrap.

The equalizer should be configured to make the low-pass equivalent
model of the system as close as possible to the ideal scheme considered
in Fig.~\ref{fig01}. Considering that the amplified-spontaneous-emission
(ASE) noise accumulated during propagation can be modeled, with good
approximation, as independent AWGN on each polarization at the input
(or, equivalently, at the output) of the fiber link, and that the
transfer matrix of the link $\mathbf{H}_{f}(f)$ (in the linear regime)
is unitary, i.e, $\mathbf{H}_{f}(f)^{-1}=\mathbf{H}_{f}(f)^{\dagger}$,
the required transfer matrix of the 2D-FFE equalizer should be%
\footnote{In the following, $(\cdot)^{*}$ denotes conjugate and $(\cdot)^{\dagger}$
transpose conjugate. For the sake of simplicity, we assume that the
analog components at transmitter and receiver are perfectly balanced
among the four tributaries, such that they can be modeled by scalar
transfer functions. Moreover, we neglect nonlinear effects, polarization
dependent loss (PDL), and optical filters along the optical link,
such that fiber propagation can be modeled by a unitary transfer matrix.
In deriving the algorithm described later, some of these assumptions
could be safely removed (e.g., including in-line optical filters),
as the proposed approach works even in the presence of colored noise
\cite{Ung:TCOM74}. On the other hand, in the presence of PDL, hardware
imbalances, and nonlinear effects, the optimality of the proposed
approach should be carefully investigated and a separate processing
of the in-phase and quadrature components (by a four-dimensional real-coefficient
equalizer) would be required. Nevertheless, as shown by the experimental
results of Section~\ref{sec:Experimental-Demonstration}, the proposed
algorithm still works satisfactorily in the presence of the aforementioned
effects (which are naturally present in the experimental setup).%
} 
\begin{equation}
\mathbf{H}_{\mathrm{eq}}(f)=\mathbf{H}_{f}(f)^{\dagger}P(f)^{*}/H_{\mathrm{fe}}(f)\label{eq:equalizer_transfer_matrix}
\end{equation}
where $H_{\mathrm{fe}}(f)$ is the low-pass equivalent transfer function
of the optoelectronic front-end (optical filter, photodetector, and
A/D converter). In this case, the corresponding overall channel transfer
matrix would be $\mathbf{H}(f)=|P(f)|^{2}\mathbf{I}$, with $\mathbf{I}$
the $2\times2$ identity matrix, independently of the actual transfer
matrix of the fiber. A naive evaluation of (\ref{eq:equalizer_transfer_matrix})
for the system at hand requires an accurate characterization of transmitter
and receiver front-end, and an adaptive estimate of the fiber transfer
matrix $\mathbf{H}_{f}(f)$. Here, instead, taking inspiration from
\cite{Ung:TCOM74}, an algorithm has been devised that configures
the equalizer according to (\ref{eq:equalizer_transfer_matrix}),
without requiring a separate knowledge of $\mathbf{H}_{f}(f)$, $P(f)$,
and $H_{\mathrm{fe}}(f)$. Denoting by $\mathbf{C}_{i}$ the $2\times2$
matrix of coefficients of the $i$-th tap of the equalizer, the equalized
samples are
\begin{equation}
\mathbf{z}_{k}=\sum_{i=0}^{N_{c}-1}\mathbf{C}_{i}\mathbf{s}_{k-i}
\end{equation}
Denoting by $\mathbf{x}_{k}$ the $k$-th column vector of transmitted
symbols, by $\mathbf{h}_{i}$ the column vector of the two $i$-th
coefficients of the desired (but unknown) overall impulse responses
at the output of the matched filter (one per polarization), and by
$\mathbf{g}_{k}=(e^{j\hat{\theta}_{1,k}},e^{j\hat{\theta}_{2,k}})^{T}$
the column vector of the phase estimates for the $k$-th samples on
the two polarizations, the error with respect to the desired channel
response is
\begin{equation}
\mathbf{e}_{k}=\mathbf{g}_{k}^{*}\circ\mathbf{z}_{k}-\sum_{i=-L_{T}}^{L_{T}}\mathbf{h}_{i}\circ\mathbf{x}_{k-i}\label{eq:error_samples}
\end{equation}
where $\circ$ denotes the Hadamard (entrywise) product and $L_{T}$
is the design parameter introduced in Section~\ref{sec:System-Design}.
As shown in \cite{Ung:TCOM74}, the variance of each element of (\ref{eq:error_samples})
is minimum when the matched filter condition is met, i.e., when $\mathbf{H}_{\mathrm{eq}}(f)H_{\mathrm{fe}}(f)\mathbf{H}_{f}(f)=P^{*}(f)\mathbf{I}$.
Given the unitarity of $\mathbf{H}_{f}(f)$, this is equivalent to
(\ref{eq:equalizer_transfer_matrix}) and provides the desired overall
response $\mathbf{H}(f)=|P(f)|^{2}\mathbf{I}$. Thus, both the required
equalizer coefficients and the desired channel coefficients of the
Ungerboeck observation model can be simultaneously estimated by an
iterative data-aided stochastic-gradient algorithm that minimizes
the variance of (\ref{eq:error_samples}). By holding $\mathbf{h}_{0}$
constant (to an arbitrary value) and forcing the symmetry condition
$\mathbf{h}_{-i}=\mathbf{h}_{i}^{*}$, the update law for the equalizer
coefficients and the estimated channel coefficients are, respectively
\begin{align}
\mathbf{C}_{i}^{(k+1)} & =\mathbf{C}_{i}^{(k)}-\alpha_{c}(\mathbf{g}_{k}\circ\mathbf{e}_{k})\mathbf{s}_{k-i}^{\dagger},\quad\quad0\le i\le N_{c}-1\label{eq:equal.coeff.updates}\\
\mathbf{h}_{i}^{(k+1)} & =\mathbf{h}_{i}^{(k)}+\alpha_{g}(\mathbf{e}_{k}\circ\mathbf{x}_{k-i}^{*}+\mathbf{e}_{k}^{*}\circ\mathbf{x}_{k+i}),\quad1\le i\le L_{T}\label{eq:channel_coeff.updates}
\end{align}
where $\alpha_{c}$ and $\alpha_{h}$ are the step-size gains. For
an accurate estimate of the whole channel response, $L_{T}$ should
be taken at least equal to $L$, where $2L+1$ is the overall channel
memory. However, by choosing a lower $L_{T}$, the equalizer forces
ISI to zero beyond the previous and past $L_{T}$ symbols, which provides
a better performance when combining the equalizer with the $2^{L_{T}}$-state
BCJR detectors \cite{Ung:TCOM74}. Updates (\ref{eq:equal.coeff.updates})
and (\ref{eq:channel_coeff.updates}) require knowledge of the transmitted
symbols. While the equalizer coefficients need to be continuously
updated to track variations of the fiber-optic channel, coefficients
$\{\mathbf{h}_{i}\}$ of the overall channel response do not change
with time and can be estimated only once when setting up the link,
unless the link transfer matrix is not unitary (e.g., due to PDL).
The initial convergence of the algorithm can be guaranteed by the
use of a known training sequence, while a slow tracking of the fiber
channel can be achieved by updating only the equalizer coefficients
according to (\ref{eq:equal.coeff.updates}), possibly at a much lower
rate than $1/T$ and with a significant delay%
\footnote{As variations of the fiber-optic channel typically take place on a
time scale of milliseconds, the channel remains approximately constant
over many consecutive codewords.%
}. This allows to use pilot symbols and/or to replace transmitted symbols
with final decisions (after successful decoding of the whole codeword),
with a negligible impact on information rate and performance.

The computation of the channel metrics for the BCJR algorithm requires
knowledge of the channel coefficients $\{\mathbf{h}_{i}\}$ and of
the noise variance. Thus, once estimated by (\ref{eq:channel_coeff.updates}),
channel coefficients are passed to the BCJR processing blocks together
with an estimate of the variance of (\ref{eq:error_samples}).

\section{Experimental Demonstration\label{sec:Experimental-Demonstration}}

\begin{figure*}
\begin{centering}
\includegraphics[width=1\textwidth]{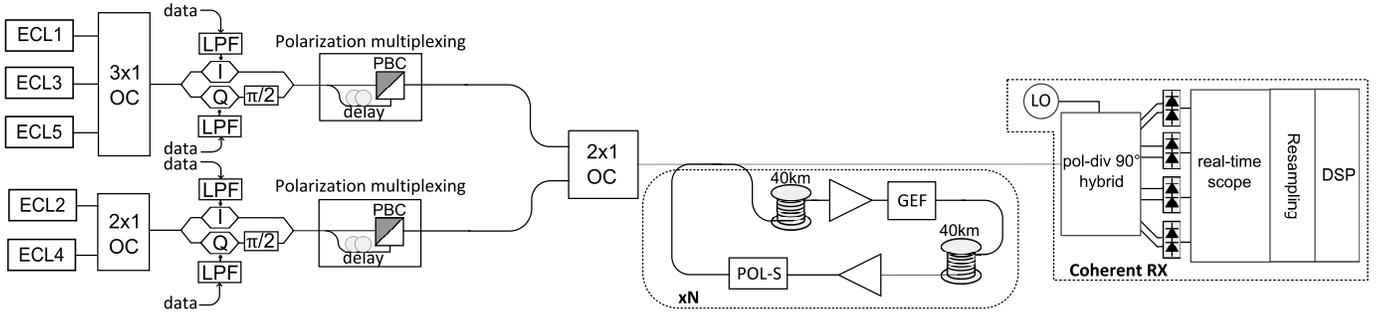}
\par\end{centering}

\caption{\label{fig06}Experimental setup.}
\end{figure*}

\begin{figure}
\centering{}\includegraphics[width=1\columnwidth]{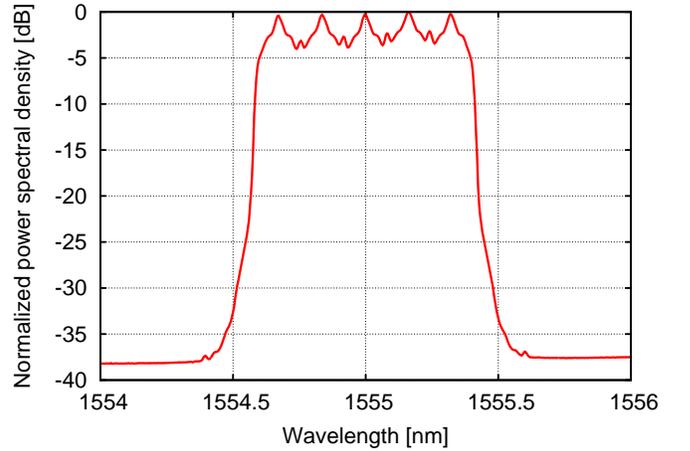}\caption{\label{fig07}Normalized optical spectrum of the TFP superchannel
at the input of the recirculating loop.}
\end{figure}

Fig.~\ref{fig06} shows the experimental setup employed for the practical
implementation of the TFP system and for the transmission experiments.
 Five external-cavity lasers (ECL) are grouped into two sets (odd
and even channels), which are separately modulated by means of two
integrated nested Mach-Zehnder modulators (IQ-MZM). Bandwidth, rate,
and spacing of the five TFP channels are optimized (under some constraints
posed by the available hardware) according to the design procedure
described in Section~\ref{sec:System-Design} to maximize the achievable
SE with the desired detector complexity. In particular, the optical
carrier spacing is set to $F=\unit[20]{GHz}$ and the binary electrical
signals that drive the in-phase (I) and quadrature (Q) port of each
IQ-MZM are modulated at a rate $R=1/T=\unit[40]{GBd}$ and filtered
by a ninth-order Chebyshev low-pass filter (LPF) with a 3dB-bandwidth
$B=\unit[10]{GHz}$%
\footnote{This configuration is twice more ``packed'' than allowed by the
Nyquist limit: its time-frequency spacing product is $FT=0.5$, meaning
that two QPSK symbols are transmitted per second per Hertz per polarization.%
}. The peak-to-peak modulation voltage is set to $V_{\mathrm{pp}}=\unit[1.5]{V}$,
while the half-wave voltage of each MZM is \foreignlanguage{english}{$V_{\pi}=\unit[2.8]{V}$.}
For the same fixed bandwidth and spacing, lower transmission rates
of 35 and 30\,GBd are also considered%
\footnote{If the bandwidth of the terminal filters is not proportionally reduced,
the LDPC codes are slightly suboptimal.%
}. Polarization multiplexing is emulated by means of a 50/50 beam splitter,
an optical delay line, and a polarization beam combiner (PBC). Each
I and Q component is modulated by a sequence of random information
bits, which are independently encoded according to one of the LDPC
codes reported in Tab.~\ref{tab:parameters-of-the}. Odd and even
channels are then combined by means of a $2\times1$ optical coupler
(OC). The optical spectrum of the transmitted TFP superchannel (at
the input of the recirculating loop) is depicted in Fig.\,\ref{fig07}. 

At the receiver side, one of the five TFP channels is detected by
employing coherent phase- and polarization-diversity detection and
setting the local oscillator (LO) at the nominal wavelength of the
selected channel. The received optical signal is mixed with the LO
through a polarization-diversity $90\text{\textdegree}$ hybrid optical
coupler, whose outputs are sent to four pairs of balanced photodiodes.
The four photodetected signals are sampled and digitized through a
20\,GHz 50\,GSa/s real-time oscilloscope in separate blocks of one
million samples at a time, corresponding to about 12 codewords (at
40\,GBd) per each quadrature component. After digital resampling
at rate $1/T$ (one sample per symbol), each block is processed off-line
according to the scheme of Fig.~\ref{fig05}, with $N_{c}=23$ equalizer
taps and $L_{T}=3$ (a truncated channel memory of $2L_{T}+1=7$ symbols,
which requires an 8-state BCJR detector). The first received codeword
of each block (on each quadrature component) is used as a training
sequence for the convergence of the DSP algorithms (initial estimate
of the frequency offset $\hat{f}_{0}$, equalizer coefficients $\mathbf{C}_{i}$,
and channel coefficients $\mathbf{h}_{i}$), while the others are
effectively employed to measure system performance. This is not considered
in the computation of the SE as, in a real system, the training sequence
would be transmitted only once. After decoding of each codeword, the
equalizer coefficients are then slowly updated (one update each 500
decoded symbols) according to (\ref{eq:equal.coeff.updates}) by employing
decisions. Pilot symbols at rate $r_{p}=1/400$ are finally employed
(and accounted for in the SE computation) to make the iterative decoding
process (CPE, BCJR detection, and LDPC decoding) bootstrap. A maximum
of 20 turbo iterations for the CPE, BCJR, and LDPC decoder is considered.

Bit-error rate (BER) measurements are performed off-line by averaging
over a total of about 500 randomly selected codewords (of length
64800). Although this result should set a limit for BER measurements
at about $10^{-6}$, it is known that LDPC error floors entails bit
errors in bursts, preventing the reliable measurement of such low
BER. However, the design of LDPC codes with low error floors falls
outside the scope of our work and we conjecture, based on extensive
computer simulations and measurement campaigns \cite{Poti:arxiv15_fieldtrialTFP},
that the employed codes do not present floors for packet error rate
higher than $10^{-7}$. This sets a limit for the minimum measurable
BER at about $10^{-7}$, and for reliable BER measurements at about
$10^{-6}$. Transmission is considered to be error-free when all the
information bits are correctly decoded at the receiver, which means,
in fact, $\mathrm{BER}<10^{-6}$ with high probability. As discussed
in Section~\ref{sec:System-Design}, detection of the adopted LDPC
codes is characterized by the presence of possible error floors at
$\mathrm{BER}<10^{-8}$ (a too low value to be observed either experimentally
or in standard simulations), which can be practically removed by concatenating
outer hard-decision BCH codes with small additional overhead (<0.4\%)
and complexity \cite{Standard-DVB-S2}. We account for this fact by
virtually including a BCH code with rate $r_{\mathrm{BCH}}=0.996$
(0.4\% redundancy) in the computation of the experimentally achieved
SE, which is therefore defined as the actual amount of information
that is reliably transmitted (error-free within measurement accuracy)
per unit time and bandwidth once the overhead due to the LDPC code,
outer BCH code (virtually present), and pilot symbols is removed 
\begin{equation}
\tilde{\eta}\triangleq\frac{4r_{\mathrm{LDPC}}r_{\mathrm{BCH}}(1-r_{p})}{FT}\le\hat{\eta}_{\mathrm{max}}.\label{eq:exper.spettr.eff.}
\end{equation}
Depending on the available $E_{b}/N_{0}$ and on the presence of uncompensated
transmission impairments, different code rates $r_{\mathrm{LDPC}}$
are required to obtain reliable transmission. The transmission system
can thus be adapted to finely adjust the information rate to the channel
conditions (accumulated noise and propagation penalties) by changing
$r_{\mathrm{LDPC}}$ (selected among the values available in Tab.\,\ref{tab:parameters-of-the})
while keeping the transmission rate $1/T$ and channel spacing $F$
constant. A wider tuning of the SE is finally obtained by changing
also the transmission rate. In particular, rates of 40, 35, and 30\,GBd
are considered. The difference between the experimentally achieved
SE (\ref{eq:exper.spettr.eff.}) and the theoretically achievable
SE (\ref{eq:max.ach.spettr.eff.}) depends on the performance (and
available rate granularity) of the designed LDPC codes and on the
presence of any additional impairment unaccounted for in this design
procedure (e.g., modulator imperfections, nonlinearity, etc.).

Long-distance transmission is emulated by using a recirculating loop,
composed of two 40\,km long spans of standard single mode fiber,
two optical amplifiers, a polarization scrambler (POL-S), and a gain
equalizer filter (GEF). The total dispersion accumulated during propagation
through the recirculating loop is compensated by a static frequency-domain
equalizer, placed in front of the 2D-FFE equalizer and configured
according to the selected link length.

\subsection{Back to back measurements}

\begin{figure}
\begin{centering}
\includegraphics[width=1\columnwidth]{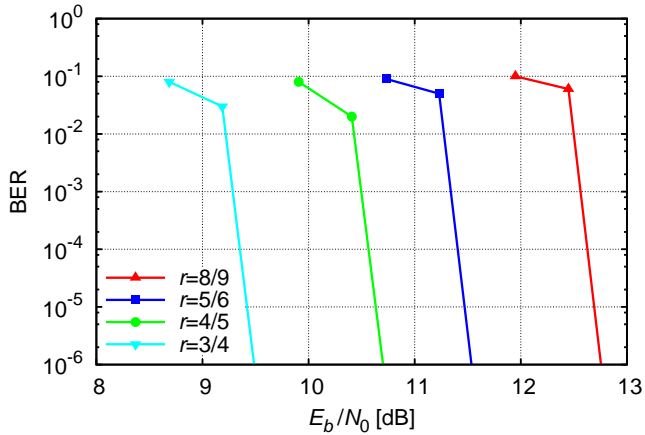}
\par\end{centering}

\caption{\label{fig08}Experimental back-to-back performance of the TFP system
(only central channel): BER with the 40\,GBd DP-QPSK configuration
and different code rates.}
\end{figure}

\begin{figure}
\begin{centering}
\includegraphics[width=1\columnwidth]{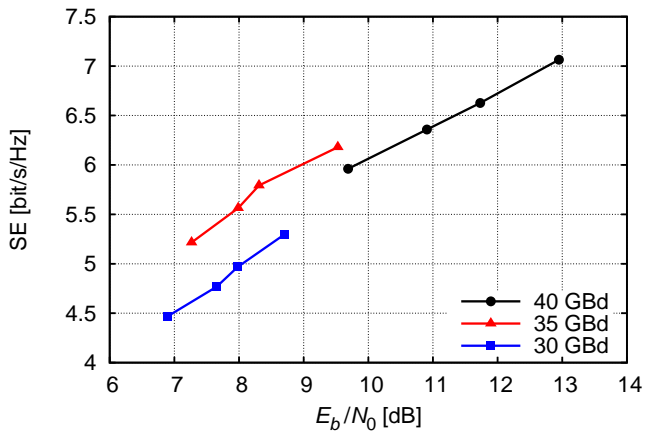}
\par\end{centering}

\caption{\label{fig09}Experimental back-to-back performance of the TFP system
(only central channel): achieved SE with the 40\,GBd, 35\,GBd, and
30\,GBd DP-QPSK configuration.}
\end{figure}

The back-to-back performance of the system is investigated by bypassing
the recirculating loop and measuring only the BER of the central channel.
In order to experimentally verify the TFP optimization performed numerically
in Section~\ref{sec:System-Design}, signals at different baud rates
are generated and coded with different code rates. Fig. \ref{fig08}
shows the measured BER values as a function of $E_{b}/N_{0}$ (obtained
by measuring the OSNR through an optical spectrum analyzer and using
the relation $E_{b}/N_{0}=2B_{\mathrm{ref}}\mathrm{OSNR}/R_{b}$)
for a 40\,GBd transmission and some of the LDPC codes reported in
Tab.\,\ref{tab:parameters-of-the}. Compared to simulation results
in Fig.\,\ref{fig03}, the experimental penalty is about 2\,dB for
the 3/4 LDPC code and increases up to more than 3\,dB for the 8/9
code. This can be explained by considering that the measured $E_{b}/N_{0}$
ratio reported on the x-axis accounts only for optical noise. Thus,
we expect an experimental penalty due to electrical noise (and other
receiver imperfections) that becomes more relevant as the measured
$E_{b}/N_{0}$ increases (i.e., the amount of optical noise decreases).
Fig.\,\ref{fig09} shows the achieved SE, defined according to (\ref{eq:exper.spettr.eff.}),
for 30, 35, and 40\,GBd transmission. In practice, Fig.\,\ref{fig09}
reports, for each code rate and baud rate, the achieved SE and the
corresponding minimum required $E_{b}/N_{0}$ ratio to obtain reliable
transmission (where the BER curves in Fig.\,\ref{fig08} suddenly
drop to zero). 
As predicted by Fig.\,\ref{fig02}, the highest SE is achieved at
40\,GBd transmission (and could be possibly increased by further
increasing the transmission rate up to 50\,GBd, though we could not
verify it due to limitations of the available hardware). However,
for low SNRs, a slightly better efficiency can be obtained at 35\,GBd.

\subsection{Transmission experiments}

\begin{figure}
\centering{}\includegraphics[width=1\columnwidth]{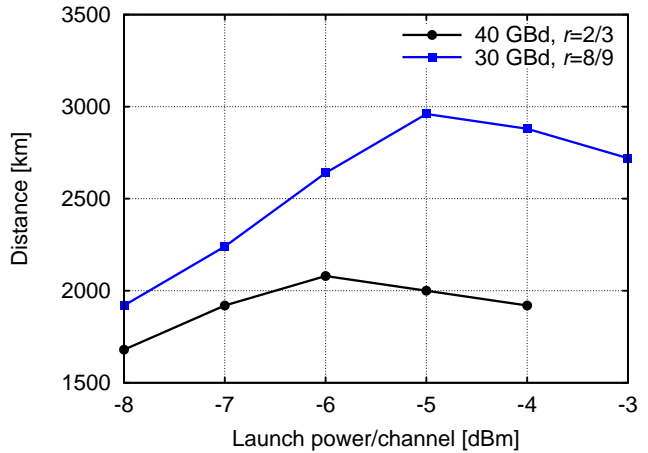}\caption{\label{fig10}Optimization of the launch power: reached distance vs
launch power at a fixed SE $\tilde{\eta}\simeq\unit[5.3]{bit/s/Hz}$.}
\end{figure}

\begin{figure}
\centering{}\includegraphics[width=1\columnwidth]{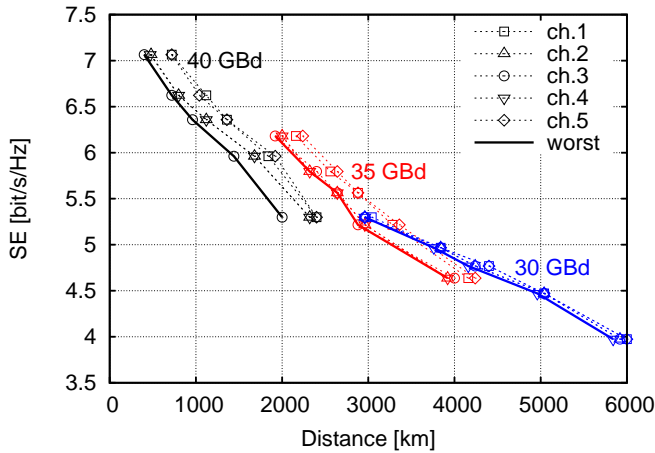}\caption{\label{fig11}Experimentally achieved SE (all the TFP channels) vs
reached distance with the 40\,GBd, 35\,GBd, and 30\,GBd DP-QPSK
configuration.}
\end{figure}

Transmission experiments are performed by properly setting the number
of rounds that the signal travels through the recirculating loop in
Fig.~\ref{fig06}. The launch power is optimized to obtain the best
trade-off between noise and nonlinear propagation effects. For the
sake of simplicity, it is assumed that the optimal launch power is
independent of the transmission distance and code rate. The optimization
is performed by setting the same power for the five channels and measuring
the performance of the third (central) one, which is the most affected
by inter-channel nonlinearity. Fig.\,\ref{fig10} shows the maximum
achievable transmission distance as a function of the launch power,
for either 30 or 40\,GBd transmission and a fixed SE $\tilde{\eta}\simeq\unit[5.3]{bit/s/Hz}$
(obtained with code rates of 8/9 and 2/3, respectively). A slightly
different result is obtained for 30\,GBd and 40\,GBd transmission,
the optimum launch power being -5 and -6\,dBm per channel, respectively.
However, in the following measurements, the same launch power of -5\,dBm
per channel is used for any transmission rate. Once the launch power
has been set, the maximum achieved SE $\tilde{\eta}$---defined in
(\ref{eq:exper.spettr.eff.}) and obtained by selecting the highest
code rate (among the available ones reported in Tab.\,\ref{tab:parameters-of-the})
guaranteeing error-free transmission---is measured as a function of
the transmission distance for a transmission rate of 30, 35, and 40\,GBd.
Fig.\,\ref{fig11} shows the results for each of the five TFP channels
(symbols) as well as for the whole super-channel (lines, corresponding
to the worst-performing channel). Due to inter-channel nonlinearity
(as $F=2B$, linear crosstalk among channels is practically negligible),
the central channel is typically the worst performing, while the outer
channels are the best performing. This is more evident at higher transmission
rates. At short distances, i.e. at high OSNRs, the achieved SE is
much higher (about 7.1\,bit/s/Hz at 400\,km) than the theoretical
limit of 4\,bit/s/Hz achievable by Nyquist-WDM transmission with
same DP-QPSK modulation format, and remains higher up to almost 6000\,km.
Moreover, the SE can be adapted to the propagation conditions by simply
changing the code rate or, for significant OSNR variations, the amount
of packing (i.e., the baud rate $1/T$ or, equivalently, the bandwidth
$B$ and the frequency spacing $F$), without changing the modulation
format and transceiver hardware.

\section{Discussion\label{sec:Discussion}}

\begin{figure}
\begin{centering}
\includegraphics[width=1\columnwidth]{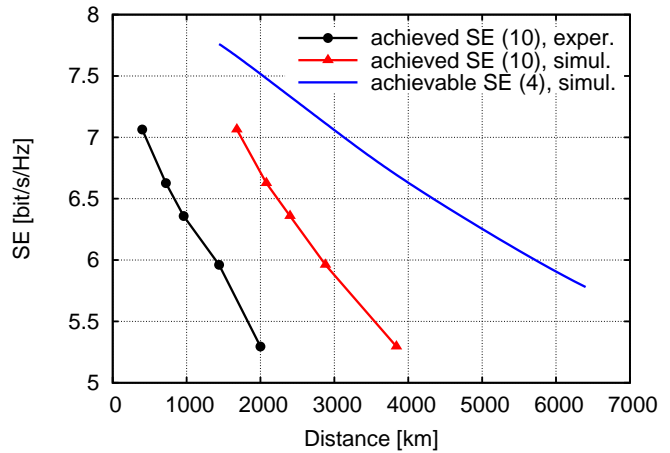}
\par\end{centering}

\caption{\label{fig:confronto-experimento-simulazioni}Comparison among the
SE achieved with the experimental setup, the SE achieved in numerical
simulations, and the theoretically achievable SE.}
\end{figure}

As a recent and not yet fully mature technique, TFP can be still improved
in terms of performance and complexity. In this sense, the implementation
proposed in this work is intended to demonstrate the technical feasibility
and good performance of TFP and should not be considered as the ultimate
TFP solution. In fact, there are several options to improve the performance
(SE vs. distance) of the proposed TFP transmission technique beyond
that achieved in our experiments. In the first place, we consider
some improvements that, being related to the design, optimization,
or implementation of the system, do not affect its complexity. As
indicated in Fig.\,\ref{fig02}, due to some limitations in the available
hardware, the adopted configuration is not exactly the optimum one.
In fact, according to numerical simulations, the optimum configuration
provides from 0.5 to 1\,bit/s/Hz of improvement of the achievable
SE (for low and high OSNRs, respectively) compared to the suboptimum
configuration actually employed in the experimental setup. Moreover,
as the optimization of Fig.\,\ref{fig02} refers to an AWGN channel,
a more accurate optimization could be sought by accounting also for
nonlinear effects through approximate channel models, time-consuming
simulations, or directly optimizing the experimental setup through
extensive measurements (in increasing order of accuracy and required
time). A non-negligible improvement should be achievable, still without
increasing the system complexity, also by removing possible imperfections
in the experimental setup (e.g, unbalance between I/Q components or
polarizations at the modulator or at the optoelectronic front-end).
This could provide up to 2\,dB of improvement in the required SNR,
as suggested by the comparison between Fig.\,\ref{fig03} and \ref{fig08}.
A similar improvement could be achieved also by designing better LDPC
codes (with a lower distance from the Shannon limit), as discussed
in Section\,\ref{sec:System-Design}. Such an improvement in the
required SNR would then translate into an almost proportional improvement
in terms of maximum transmission distance. This is shown in Fig.\,\ref{fig:confronto-experimento-simulazioni},
which compares the achieved SE (\ref{eq:exper.spettr.eff.}) measured
in the experimental setup (the 40\,GBd configuration), the achieved
SE (\ref{eq:exper.spettr.eff.}) estimated through numerical simulations
(including nonlinear effects but without accounting for TX or RX imperfections),
and the achievable SE (\ref{eq:max.ach.spettr.eff.}) (for the best
possible code) over the nonlinear optical channel. The latter is estimated
through numerical simulations as in \cite{Colavolpe2014_TCOM_TFPacking},
in which the output samples $\mathbf{y}^{(0)}$ in (\ref{eq:AIR})
are obtained by propagating the input samples $\mathbf{x}^{(0)}$
through the nonlinear optical fiber link by means of the SSFM algorithm,
such that the expectation in (\ref{eq:AIR}) is evaluated according
to the actual sample distribution over the nonlinear channel and represents
the AIR over that channel. The difference between the first two curves
(experiments and simulations) is due to the aforementioned imperfections
of the experimental setup, which are not considered in the numerical
simulations and could be possibly removed. Moreover, for high OSNR
values (i.e., at short distances), electronic noise and quantization
effects at the receiver (not included in the simulations) become relevant
with respect to optical noise and causes an additional penalty with
respect to numerical simulations. This explains why the difference
between the two curves (in dBs along the x-axis) is not constant and
increases at short distances. Finally, the difference between the
last two curves (achieved and achievable SE) depends on the actual
performance of the designed LDPC codes compared to the ultimate limit
provided by the information theoretical analysis.

In the second place, a significant improvement can be obtained at
the expense of an additional DSP complexity. For instance, by modifying
the detection strategy to account for a longer ISI (e.g, increasing
the number of trellis states or considering channel shortening techniques
\cite{Rusek2012_TWC_Channel-shortening}, as described in \cite{Colavolpe2014_TCOM_TFPacking})
and/or also for ICI (multi-user detection), pulses can be more densely
packed, achieving a higher SE. Finally, the DSP implemented in the
experimental setup does not include any nonlinearity mitigation strategy,
which could be adopted to improve the overall performance. For instance,
as confirmed by some preliminary results, the low-complexity digital
backpropagation strategy proposed in \cite{Secondini2014_ECOC_ESSFM}
can be easily integrated in the DSP (replacing the static frequency-domain
equalizer for dispersion compensation) to mitigate intra-channel nonlinearity
and extend the reach. This subject is however outside the scope of
the paper and is left for a future investigation.

Due to possible improvements and ongoing research, an accurate and
comprehensive comparison with more conventional (and mature) techniques
(e.g., Nyquist-WDM) is not yet available. A numerical comparison between
TFP and Nyquist-WDM performance can be found in \cite{Colavolpe2014_TCOM_TFPacking},
where it is shown that TFP can achieve higher SE values than high-order
modulation Nyquist-WDM over realistic long-haul systems. In terms
of peak-to-average power ratio (PAR), it is known that non-orthogonal
signaling (FTN or TFP) has a higher PAR than orthogonal signaling
employing the same (QPSK) symbol alphabet, in effect inducing a sort
of ``higher-order constellation'' which foils the beneficial effect
that a constant-envelope modulation has in terms of robustness to
nonlinear effects. The effect of PAR on nonlinearity becomes, however,
almost negligible in dispersion-unmanaged systems with relevant span
losses \cite{Sec:PTL14}. Moreover, as shown in \cite{Colavolpe2014_TCOM_TFPacking},
TFP outperforms high-order Nyquist-WDM especially in the presence
of nonlinear effects. The actual PAR is not relevant even in terms
of the resolution of the required A/D or D/A converters. In fact,
no D/A is employed to generate the TFP signal. On the other hand,
received samples in dispersion-unmanaged systems are always (with
high accuracy) Gaussian distributed because of the accumulated dispersion
\cite{carena2010statistical}, such that the PAR of the received signal
is independent of the modulation format. TFP offers advantages also
at a network level, as it provides high SE and flexibility (e.g.,
reach adaptation and filter configuration) without requiring transponders
supporting multiple modulation formats \cite{Sambo2014_JLT_TFP}.
In terms of complexity, TFP has the advantage of a simpler transmitter
architecture (e.g., only two-level driving signals are needed to control
the modulator; no DSP and D/A conversion are required to process the
modulating signals; more relaxed constraints on the pulse shape can
be considered) at the expense of a more complex DSP at the receiver
(e.g., the 8-state BCJR detector employed in this work and the higher
symbol rate that affects the rate of operations that algorithms operating
at symbol time should perform). On the other hand, the use of a TFP
DP-QPSK format allows to employ only symbol-time processing at the
receiver and greatly simplifies decision-directed algorithms compared
to higher-order modulation formats. This can partly compensate for
the additional complexity of the BCJR detector.

\section{Conclusions\label{sec:Conclusions}}

In this work, after reviewing the main theoretical aspects of the
TFP approach, we have investigated its application to fiber-optic
systems. The main challenges pertain to the peculiar nature of the
channel (the optical fiber, impaired by linear and nonlinear propagation
effects) and to the high data rates involved. We have thus discussed
the implementation schemes, focusing on the main differences with
respect to a conventional coherent WDM system. The only relevant difference
is a modification of the DSP algorithms employed for detection. In
the proposed scheme, a butterfly equalizer adaptively addresses propagation
impairments and performs matched filtering, while intentional ISI
due to TFP filtering is accounted for by a BCJR detector. To ensure
a proper distribution of tasks between the equalizer and BCJR detector,
an algorithm has been proposed that adaptively controls the equalizer
and provide channel coefficients and noise variance to the BCJR detector.
This makes the receiver fully adaptive, without requiring a priori
knowledge of the adopted TFP configuration. Soft-decoding forward-error
correction is finally employed. In particular, irregular LDPC codes
with various code rates (in the range 2/3--8/9) and specifically optimized
for the TFP channel have been designed to operate at low error rates.
They approach (within about 3\,dB) the SE limits achievable by the
proposed techniques at different SNRs. The performance of the proposed
system has been tested both experimentally and by simulations, demonstrating
technical feasibility and good performance. Five closely-packed DP-QPSK
channels were transmitted through a recirculating loop, keeping the
net SE beyond the theoretical limit of Nyquist-WDM (4\,bit/s/Hz)
up to 6000\,km. The channel bandwidth and spacing was held fixed
to 20\,GHz, while the transmission rate and code rate were adapted,
depending on the transmission distance, to the available OSNR. At
400\,km, a net SE of more than 7\,bit/s/Hz was achieved by setting
the transmission rate at 40\,GBd (twice faster than the Nyquist limit)
and the code rate to 8/9. The transmission distance was then gradually
increased up to 6000\,km, with a net SE which gradually decreased
to about 4\,bit/s/Hz (achieved with a 30\,GBd transmission rate
and a 2/3 code rate). Both the LDPC encoder and decoder were actually
included in the experimental setup, as it is advisable in the presence
of soft decoding, for which the use of a numerically evaluated ``pre-FEC
BER threshold'' may be unreliable. 

In conclusion, we have demonstrated that TFP with low-level modulation
(e.g., DP-QPSK) can be considered as a practical and viable alternative
to high-level modulations to achieve high SEs over long-haul fiber-optic
links, providing good performance and high flexibility (reach adaptation
and filter configuration) with simpler transponder architectures (single
modulation format, relaxed constraints on pulse shape, no DSP and
D/A conversion at the transmitter).

\selectlanguage{english}%
\vspace*{2.3ex}
\IEEEtriggeratref{26}
\selectlanguage{american}%

\end{document}